\documentclass{INTERSPEECH2023}
\usepackage{hyperref}
\usepackage{adjustbox}
\usepackage[noend]{algpseudocode}
\usepackage{algorithm}
\usepackage{algpseudocode}
\usepackage{amsmath}
\usepackage{graphics}
\usepackage{epsfig}
\usepackage[table,xcdraw]{xcolor}
\usepackage{multirow}
\usepackage{makecell}

\interspeechcameraready

\title{Parameter-Efficient Learning for Text-to-Speech Accent Adaptation}
\name{Li-Jen Yang$^\star$,	Chao-Han Huck Yang$^\dagger$, 	Jen-Tzung 	Chien$^\star$}
\address{
  $^\star$National Yang Ming Chiao Tung University, Taiwan\\
  $^\dagger$Georgia Institute of Technology, USA }
\email{lijen0918.ee10@nycu.edu.tw, 	huckiyang@gatech.edu, jtchien@nycu.edu.tw}

\begin{document}

\maketitle
 
\begin{abstract}
This paper presents a parameter-efficient learning (PEL) to develop a low-resource accent adaptation for text-to-speech (TTS). A resource-efficient adaptation from a frozen pre-trained TTS model is developed by using only 1.2\% to 0.8\% of original trainable parameters to achieve competitive performance in voice synthesis. Motivated by a theoretical foundation of optimal transport (OT), this study carries out PEL for TTS where an auxiliary unsupervised loss based on OT is introduced to maximize a difference between the pre-trained source domain and the (unseen) target domain, in addition to its supervised training loss. Further, we leverage upon this unsupervised loss refinement to boost system performance via either sliced Wasserstein distance or maximum mean discrepancy. The merit of this work is demonstrated by fulfilling PEL solutions based on residual adapter learning, and model reprogramming when evaluating the Mandarin accent adaptation. Experiment results show that the proposed methods can achieve competitive naturalness with parameter-efficient decoder fine-tuning, and the auxiliary unsupervised loss improves model performance empirically.
\end{abstract}

\section{Introduction}
Large-scaled pre-trained acoustic models~\cite{zhang2023google, zhang2022bigssl} and language models~\cite{brown2020language} or so-called foundation models~\cite{bommasani2021opportunities} have been emerging due to the rapid development of efficient computation hardware and self-supervised learning. Recently, the generative diffusion models \cite{popov2021grad,rombach2022high, ho2020denoising} have achieved dominant performance across different tasks. Either pre-trained foundation models or generative diffusion models require powerful computation hardware and long training time. Therefore, parameter-efficient adaptation plays an important role in many practical applications when utilizing the pre-trained model for a low-resource downstream task. The approaches such as prompt tuning \cite{lester2021power}, residual adapter \cite{rebuffi2017learning} or model reprogramming \cite{elsayedadversarial, yang2021voice2series, hambardzumyan2021warp} have been developed for parameter-efficient learning (PEL)~\cite{hetowards} where several benefits are commonly pursued. First, the training time is reduced by freezing the backbone model and only adapting the domain-specific parameters. Second, the generalization to a low-resource and out-of-distribution target domain is improved. In \cite{yang2021voice2series, kumar2022fine}, fine-tuning the whole model was seen as a way to distort the pre-trained features for an out-of-distribution target task. The way of updating a small portion of a pre-trained model or freezing an entire backbone model and updating only the extra weights was illustrated as the lightweight fine-tuning which generalized the representation under distribution shift. Third, the backbone model is reused \cite{lester2021power,hung2023low} for those methods which only add on additional weights, arrange task-specific prompts or reprogram input layers for different tasks. Several works were recently proposed to introduce the adapter learning and model reprogramming in speech-related applications. In \cite{yang2023english}, a pre-trained English automatic speech recognition (ASR) model was repurposed as a multilingual ASR by adding reprogramming layers. In \cite{morioka2022residual}, residual adapters were added to a backbone model for speaker adaptation in a text-to-speech (TTS), ensuring the naturalness of synthesized speech through the use of only a few controllable parameters.

Meanwhile, accent adaptation~\cite{chien1999extraction, saon2012large, yamagishi2010thousands} is seen as a practical issue for TTS system which is handled to develop the generative model for speech synthesis in presence of pronunciation variations or accent shifts due to different speakers in different ages coming from different regions. For instance, accent adaptation is one essential step when conducting the English accent transfer among different countries including United States, United Kingdom and Australia. For Chinese spoken language, similar challenges~\cite{ji2004culture} have been reported due to differences in hearing effects between Mainland Chinese accent (zh-CN) and Taiwanese Mandarin accent (zh-TW) on voice quality and understanding. One important task in TTS is to adapt additional layers and weights for accent adaptation based on the frozen pre-trained TTS backbone. This study addresses the emerging challenges~\cite{ji2004culture} of data sparseness for Taiwanese Mandarin accent in accent adaptation by utilizing a pre-trained TTS model trained on a large corpus of Chinese speech with Mainland China accent. A direct way to tackle this problem is to fine-tune the whole pre-trained model \cite{li-iscslp}, but the computation cost is still high due to a large number of parameters. Accordingly, this paper presents and investigates several parameter-efficient methods to deal with accent adaptation for low-resource spoken language. The first study on model reprogramming scheme to TTS is explored. An additional scheme of learning residual adapter~\cite{tomanek2021residual, morioka2022residual} for speaker adaptation in TTS is evaluated. The main contributions of this paper are summarized as follows. First, two PEL methods to accent adaptation by utilizing pre-trained TTS model were presented. Second, a novel PEL based on model reprogramming for accent adaptation is proposed. The input-based reprogramming for TTS is exploited and can be applied for model re-deployment~\cite{xiao2023offsite}. Third, a model regularization based on optimal transport is developed to improve TTS performance by characterizing the latent feature distance.


\section{Background Survey}

\subsection{TTS by Utilizing Pre-Trained Model}
Many prior works on speaker adaptation~\cite{morioka2022residual, chien1999extraction} or accent adaptation~\cite{li-iscslp} were developed by conducting knowledge transfer based on a pretrained multi-speaker TTS model. A simple method is to fine-tune or adapt the entire pretrained model to a target speaker. This method is feasible to synthesize the speech signals with high naturalness and speaker characteristics. However, the drawback~\cite{hetowards} is the scaling issue which causes high computation for adapting so many parameters. In \cite{chenadaspeech}, AdaSpeech was proposed as a parameter-efficient TTS for a new speaker by additionally performing the conditional layer normalization. In \cite{morioka2022residual}, the residual adapter scheme was shown as an effective approach to speaker adaptation by adapting the prosody features of a TTS model from source to target domains through the adapter layers. This study addresses the parameter-efficient methods, e.g. adapter learning, for accent adaptation and further explores a new PEL method based on model reprogramming where the Chinese accent adaptation is implemented.
\subsection{Parameter-Efficient Learning}
There are three major approaches to parameter-efficient learning which are input prompting, adapter learning and model reprogramming. In \cite{brown2020language}, the generative pre-trained transformer-3 (GPT-3), was proposed by adopting the prompt and guiding the learned model to generate the related response. In \cite{ lester2021power, gu2021ppt}, the bidirectional encoder representations from transformers (BERT) was utilized to perform domain adaptation. By concatenating the trainable task-specific prompts with input text sequence, the gap between pre-training tasks and downstream tasks was bridged. In \cite{morioka2022residual,houlsby2019parameter, sung2022vl}, the sub-module of transformer layer using adapter or residual adapter was added and adjusted so as to obtain remarkable performance for a target task. In addition, the model reprogramming or adversarial reprogramming was proposed to reprogram input data by introducing trainable layers which attempted to translate new inputs into source domain so that the pre-trained model could be used directly. For example, in \cite{yang2021voice2series}, Voice2Series was proposed to transform an input time series so that a large acoustic model was utilized to obtain competitive results on various series classification tasks. This paper is motivated to develop various parameter-efficient methods for accent adaptation with low-resource setting in a TTS system.

% ...
\label{section:doubleblind}
\label{section:preprints}

\section{Low-Resource Accent Adaptation}
\begin{figure} \centering
    \includegraphics[width=.8\columnwidth]{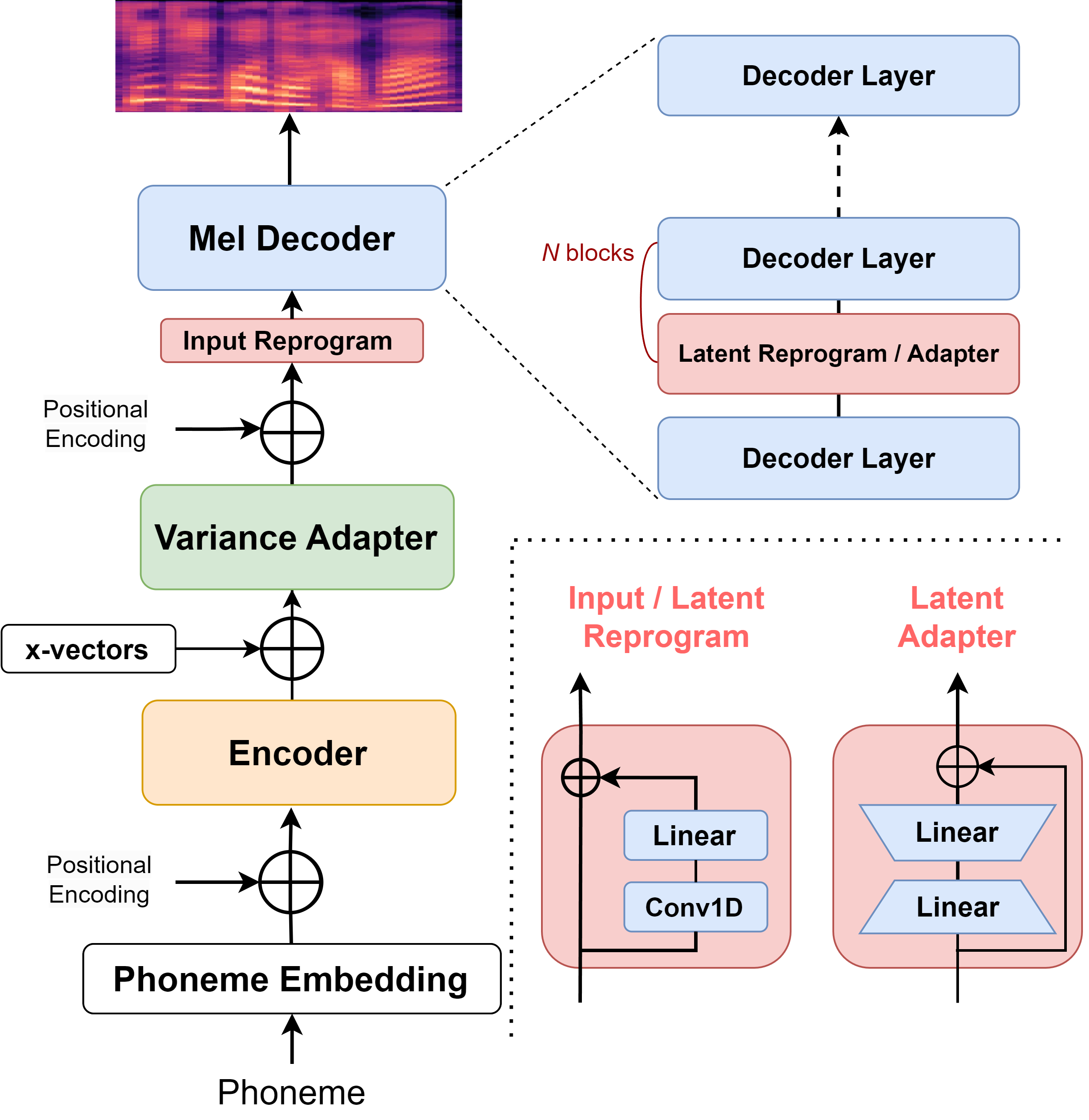}
    \caption{\label{fig:model}System architecture for parameter-efficient learning by using conformer-fastspeech2 backbone. Three kinds of layers are configured in input and latent spaces of Mel decoder.}
   \vspace{-6mm}
\end{figure}

Considering the success of parameter-efficient learning in different tasks and domains, this paper presents the parameter-efficient learning and model regularization for low-resource accent adaptation in a TTS system where the backbone model based on a \textit{conformer-fastspeech2}~\cite{renfastspeech} model from ESPNet~\cite{watanabe2018espnet} is utilized. System architecture is shown in Figure \ref{fig:model}. The speech spectrogram from phoneme input is synthesized through a stack of components consisting of phoneme embedding, encoder, variance adapter and Mel decoder where the additional position embeddings and \textit{x-vectors}~\cite{snyder2018x} are added during stack processing. Our proposed model consists of two parts: parameter learning and parameter regularization, which will be individually addressed in the following sections. 

\subsection{Parameter-Efficient Accent Adaptation}
First, parameter-efficient learning is performed by reshaping the architecture of Mel decoder through three types of layer while the remaining components in TTS are frozen. The input space and latent space of Mel decoder are re-organized by merging with input reprogramming layer, latent adapter layer and latent reprogramming layer which are shown in Figure \ref{fig:model}. The number of controllable parameters due to the add-on layers is very limited relative to the whole model architecture. Briefly speaking, input reprogramming layer aims to reduce the cost of model re-deployment based on a kind of prompt scheme. The latent reprogramming layer is implemented as separate reprogramming layer which is appended in latent space of Mel decoder. 

\subsubsection{Input reprogramming layer}
In \cite{yang2021voice2series,chang2022exploration}, input reprogramming was first proposed to implement a prompt tuning scheme to redeploy endpoint models for speech processing tasks in recent works \cite{xiao2023offsite, hung2023low}. Traditionally, it is popular to carry out domain adaptation by fine-tuning the entire model or only a portion of model. Using the fine-tuning approach, the pre-trained model should be re-tuned every time when a new task is present. The issue of computation cost is severe when a large-scaled pre-trained model such as CLIP \cite{radford2021learning}, GPT-3 \cite{brown2020language} or Wav2Vec2 \cite{baevski2020wav2vec} is utilized. Parameter-efficient learning is required to handle this issue \cite{hung2023low,xiao2023offsite}. This study introduces the input reprogramming layer in conjunction with Mel decoder of a frozen TTS model. This treatment aims to address the challenge of re-deploying accent voices under a fixed Mel decoder. Such a composite layer is seen as a trainable feature extractor $\mathcal{H}_{\theta}$ which is stacked with a linear feedforward layer and a 1-dimensional convolutional layer. The input reprogramming function $\mathcal{R}_\theta$ with the decoder input $z$ is yielded as a residual calculation by $\mathcal{R}_\theta(z) = z +{\mathcal{H}_{\theta}(z)}=z{'}$ where $z{'}$ represents the reprogrammed input of decoder that is fed into the pre-trained endpoint model, and $\theta$ is the only trainable parameter for entire TTS model, which is updated during the back-propagation process. 

\subsubsection{Latent adapter layer}
Next, a frozen TTS backbone model is utilized by configuring the adapter layer or reprogramming layer in the latent space of Mel-scale decoder where the outputs are finally used to synthesize the speech spectrogram. Basically, the adapter layer or module is formed of one linear down layer and one linear up layer which are used to extract the bottleneck features, and the residual connection performs additive feature learning to engage knowledge of adapter input. There are $N$ blocks of decoder layer and adapter layer which are stacked to form a Mel decoder. Only the adapter layer is fine-tuned, the other layers are frozen.

\subsubsection{Latent reprogramming layer}
Another type of latent configuration of Mel decoder is to add on reprogramming layer in each block to enrich the optimization procedure of a TTS model. Given a frozen decoder layer $\mathcal{F}_{{\Theta}}^{i}$ in each block $i$, the latent features $h^i$ of $i$-the decoder layer is calculated and then fed into the latent reprogramming layer $\mathcal{R}_{\theta}^i$ or latent adapter layer $\mathcal{A}_{\theta}^i$ to perform feature reprogramming or adaptation, respectively, instead of taking latent feature $h_i$ as the input of $(i+1)$-th decoder layer 
\begin{equation} \begin{aligned} \label{eq:3}
    & \underbrace{\mathcal{R}_{\theta}^i(h^i)}_{i\text{-th latent reprogram}}\rightarrow(h^{i})' \rightarrow\underbrace{\mathcal{F}_{{\Theta}}^{i+1}((h^{i})')}_{i\text{-th frozen decoder with trainable feature}} \\
& \underbrace{\mathcal{A}_{\theta}^i(h^i)}_{i\text{-th latent adapter~~~~}}\rightarrow(h^{i})' \rightarrow\underbrace{\mathcal{F}_{{\Theta}}^{i+1}((h^{i})')}_{i\text{-th frozen decoder with trainable feature}}
\end{aligned} \end{equation}
where ${\Theta}$ represents a set of non-trainable parameters across decoder layers, and $\theta_i$ denotes the $i$-th trainable feature generator for PEL using either reprogramming layer or adapter layer.

\begin{figure} \centering
    \includegraphics[width=0.48\columnwidth]{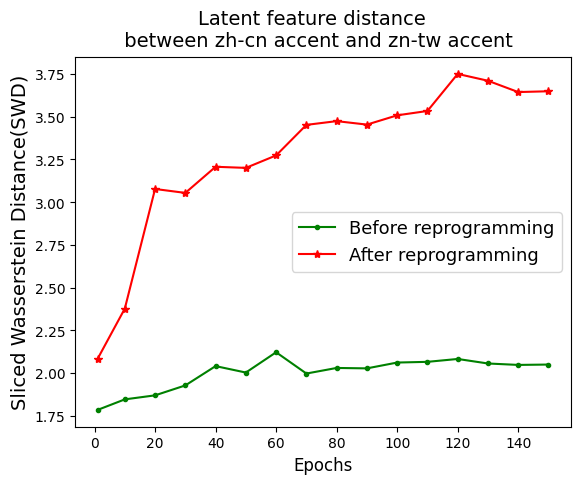}
    \includegraphics[width=0.48\columnwidth]{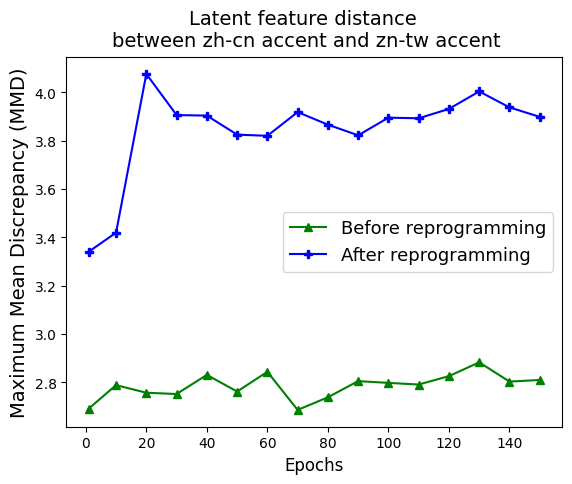}
    \caption{Latent feature distance between two accents before/after reprogramming via SWD (left) and MMD (right).}
    \label{fig:latent_distance}
    \vspace{-3mm}
\end{figure}

\subsection{Parameter Regularized Accent Adaptation}
In addition to parameter-efficient learning for accent adaptation to a low-resource target domain, this study further introduces the parameter regularization scheme in domain adaptation. In particular, an auxiliary unsupervised loss based on the optimal transport~\cite{peyre2019computational} is merged for model regularization. This consideration is based on an observation about the distance of latent features between source (Mainland Chinese) and target accents (Taiwanese Mandarin). As shown in Figure~\ref{fig:latent_distance}, the latent feature distance before and after the reprogramming layer with two metrics, sliced Wasserstein distance (SWD)~\cite{peyre2019computational} and maximum mean discrepancy (MMD)~\cite{scholkopf2002learning}, are illustrated. The latent features in accent adaptation are evaluated. Both metrics are calculated to measure the discrepancy between two probability distributions for variables $u$ and $v$. These measures belong to the family of integral probability metric (IPM)~\cite{chien2015hierarchical} which measures the optimal transport in a form of
\begin{equation} \label{eq:3}
  d_{\mathcal F} (\mu, \nu) = \sup_{f \in \mathcal F} \Big( \int f d\mu - \int f d\nu \Big)
\end{equation}
where $\mathcal F$ is a class of measurement functions. As referred in ~\cite{feydy2019interpolating}, if $f$ is selected as a $1$-Lipschitz function, SWD is a simple realization of IPM in Eq.~(\ref{eq:3}) based on Euclidean distance. If $f$ is set as a kernel function, MMD is a realization of IPM. In this evaluation, it is found that both SWD and MMD using the proposed PEL are increased after latent reprogramming along learning epochs. There is a big increase after 20 epochs. MMD converges better than SWD.

To highlight this optimal transport phenomenon during fine-tuning, a regularization term is designed and merged as auxiliary training objective to measure the distance between the source feature $h_s$ and the reprogrammed target feature $\mathcal{R}_\theta(h_t)$
\begin{equation} \begin{aligned} \label{eq:ot}
    \mathcal{L}_{\text{ot}}(h_t, h_s; \theta) = -d\left(\mathcal{R}_\theta(h_t), h_s\right)  
\end{aligned} \end{equation}
where $\mathcal{R}_\theta$ acts as either input reprogramming layer or latent reprogramming layer, and $d$ is distance metric either SWD or MMD. For the adapter, the same distance in Eq.~(\ref{eq:ot}) is used but the adaptation function $\mathcal{A}_\theta$ is adopted. As a result, the total learning objective consisting of regression loss for synthesized speech and optimal transport due to reprogramming is constructed for parameter regularized learning
\begin{equation} \begin{aligned}
   \mathcal{L} = \mathcal{L}_{\text{mae}}(\widehat{\mathbf{y}}, \mathbf{y}; \theta) + \mathcal{L}_{\text{ot}}(h_t, h_s; \theta)
\end{aligned} \end{equation}
where $\mathcal{L}_{\text{mae}}$ represents the mean absolute error (MAE) between the predicted spectrogram $\widehat{\mathbf{y}}$ from Mel decoder and the ground-truth spectrogram $\mathbf{y}$ from training speech. This MAE loss is seen as the supervised regression loss. Importantly, the optimal transport loss $\mathcal{L}_{\text{ot}}$ is calculated as the negative distance measure, namely, a penalized regularization~\cite{scholkopf2002learning} is introduced to assure separation between the source feature $h_s$ and the reprogrammed target feature $\mathcal{R}_\theta(h_t)$ or adapted target feature $\mathcal{A}_\theta(h_t)$. In the implementation, a warm-up strategy~\cite{yang2021multi} was performed by gradually increasing the coefficient of the regularization term from zero at the start of training. This treatment helps to stabilize the model. Algorithm \ref{alg:1} shows the training procedure with optimal transport as a penalized regularization.

\begin{algorithm}
  \caption{Optimal transport regularized training procedure}
  \begin{algorithmic} 
    \Require{fine-tuning data $\{\mathbf{x}, \mathbf{y}\}$, source domain feature $h_s$, total training steps $T$, hyperparameter $K$}
    \Ensure {parameter $\theta$ of PEL method} 
    \While{train~steps less than $T$}  
        \State calculate outputs $\widehat{\mathbf{y}}$ and $h_t$ given inputs $\mathbf{x}$
        \If{train~steps less than $K$}
            \State compute loss $\mathcal{L} = \mathcal{L}_{\text{mae}}(\widehat{\mathbf{y}}, \mathbf{y};\theta)$
        \Else
            \State compute loss $\mathcal{L} = \mathcal{L}_{\text{mae}}(\widehat{\mathbf{y}}, \mathbf{y};\theta) - \mathcal{L}_{\text{ot}}(h_t, h_s;\theta)$
        \EndIf
         \State update parameter $\theta$ using $\nabla_{\theta} \mathcal{L}$
    \EndWhile  
    \Return $\theta$
  \end{algorithmic}
  \label{alg:1}
 \end{algorithm}
\vspace{-3mm}

\section{Experiments}
\subsection{Experimental settings}
We chose the AISHELL3~\cite{aishell} as pretrained dataset, which consists of around 85 hours of emotion-neutral recordings delivered by 218 native Chinese mandarin speakers, to build a wide-range source accent acoustic model. Our purpose is to leverage the power of the pre-trained model and perform the accent transfer from zh-CN to zh-TW. To make sure the voice is clean and suitable for TTS, we collect a 40-minute-long Taiwanese accent corpus with only one female speaker in a quiet space. The recording transcripts are including jokes, movie introductions, and travel guides.

\begin{table}[ht]
    \centering
    \setlength{\arrayrulewidth}{0.3mm}
    \setlength\tabcolsep{3pt}
    \renewcommand{\arraystretch}{1.25}
    \caption{Objective and Subjective evaluation under different PEL methods with SWD/MMD auxiliary loss, including input reprogram (IR), latent reprogram (LR), and latent adapter (LA). }
    \footnotesize
    \begin{tabular}{|l|c|c|c|c|}
       \hline
       \textbf{Method} & \textbf{MCD} ($\downarrow$) & \makecell{\textbf{Naturalness}} ($\uparrow$) & \makecell{\textbf{AQ} ($\uparrow$)\\}  & \textbf{Params} \\
       \hline
       Ground Truth & N/A & 4.83 ± 0.38 & N/A & N/A \\
       FT  &  7.64 ± 0.76  &  4.40 ± 0.70 & 4.55 ± 0.71 &  100\%\\
       Decoder FT  &  8.06 ± 0.87  &  4.00 ± 0.95 & 4.17 ± 0.92  &   46.8\%\\
       \hline\hline
       IR & 8.07 ± 0.82 & 3.65 ± 0.85 & 3.85 ± 0.82  & \multirow{3}{*}{0.6\%} \\
       IR w/ SWD  &  8.03 ± 0.65   &   3.70 ± 0.81 & 3.98 ± 0.85  &\\
       IR w/ MMD  &  8.03 ± 0.69    &  3.70 ± 0.84 & 3.98 ± 0.88  &\\
       \hline
       LA  &   7.99 ± 0.89  & 3.77 ± 0.91 & 4.03 ± 0.76  & \multirow{3}{*}{0.8\%} \\
       LA w/ SWD  &  7.93 ± 0.78 & \cellcolor[HTML]{9AFF99}\textbf{3.88 ± 0.87} & \cellcolor[HTML]{9AFF99}\textbf{4.12 ± 0.81}  &  \\
       LA w/ MMD  &  \cellcolor[HTML]{9AFF99}\textbf{7.90 ± 0.81} &  3.73 ± 0.84 & 4.05 ± 0.86  &  \\
       \hline
       IR+LR &  7.86 ± 0.83  &  3.67 ± 0.91 & 3.92 ± 0.88  & \multirow{3}{*}{1.2\%} \\
       IR+LR w/ SWD &  7.81 ± 0.80     &  3.50 ± 0.97 & 3.90 ± 1.09 & \\
       IR+LR w/ MMD &  \cellcolor[HTML]{9AFF99}\textbf{7.79 ± 0.78}     &   \cellcolor[HTML]{9AFF99}\textbf{3.75 ± 0.80} &\cellcolor[HTML]{9AFF99}\textbf{ 3.95 ± 0.89}  &\\
    \hline
    
    \end{tabular}
    \label{tab:tw}
    \vspace{-6mm}
\end{table}

For model configuration, We employ the same pre-trained \textit{Conformer-Fastspeech2} backbone which was trained on the AISHELL3 dataset and further use the x-vector~\cite{xvectors} as speaker embedding to leverage the better speaker attribute. The entire model had 71M parameters including 4 conformer layers in both decoder and encoder which the latent feature dimension is set to 384, and it was trained using the Adam optimizer with a Transformer learning rate schedule for 500k steps. We further use the parallel-wavegan~\cite{yamamoto2020parallel} as vocoder to transform mel feature to the waveform.
For PEL methods settings, we set the bottleneck dimension size $r$ = 96 of the adapter and inserted them between 4 conformer-decoder layers, and keep the settings consistent for input reprogramming and latent reprogramming, where the Conv-1D feature extractor is set the hidden feature dimension as 96.  The training setting is same as pretraining stage but sets train steps to 20k and the step of adding auxiliary loss is set to 300.
In this work, we do the experiment on three settings: a) the input reprogram, b) the latent adapter, and c) the combination of the input reprogram and latent reprogram.
We conduct the evaluation with objective methods, mel cepstral distortion (MCD) to quantify the distortion between two sequences of Mel-frequency cepstral coefficients. We further evaluate the character error rate (CER) of pretrained Automatic Speech Recognition (ASR) model. We conduct experiments on two alternative ASR models trained with the zh-CN and zh-TW corpora of Common Voice~\cite{ardila2020common}, with CER baselines provided by HuggingFace of 0.19 and 0.10, respectively. We expect to see the synthetic Taiwanese Mandarin accent speech have lower CER on the corresponding pretrained ASR model but have a terrible result on pretrained zh-CN ASR model. Referring to the CHiVE-BERT \cite{finkelstein2022training}, we conduct a subjective mean opinion score (MOS) on naturalness and accent quality(AQ). 
The accent quality is used to determine whether the voice sample came from a native accent speaker. We ask people to rate on a 5-scale Likert scale (1: Bad, 2: Poor, 3: Fair, 4: Good, 5: Excellent) for these two MOS evaluations. 


\begin{table}[ht]
    \centering
    \setlength{\arrayrulewidth}{0.3mm}
    \renewcommand{\arraystretch}{1}
    \caption{ASR evaluation in terms of word error rate (WER) for synthetic speech was conducted under parameter-efficient settings. }
    \footnotesize
    \begin{tabular}{|l|c|c|c|}
       \hline
       \multirow{2}{*}{\textbf{Methods}} & \multicolumn{3}{c|}{\textbf{ASR Models WER ($\downarrow$)}} \\
        \cline{2-4}
        & zh-TW & zh-CN & Diff. \\
        \hline
       FT  &  0.202  & 0.408 & 0.206 \\
       Decoder FT  &  0.187 & 0.317 & 0.130 \\
       \hline\hline
       Input Reprogram & 0.210 & 0.343 &  0.133\\
       \multicolumn{1}{|r|}{w/ SWD}  &  0.215 & 0.346 & 0.115 \\
       \multicolumn{1}{|r|}{w/ MMD}  &  0.240 & 0.371 & 0.131 \\
       \hline
       Latent adapter  & 0.177 & 0.344 & 0.237\\
       \multicolumn{1}{|r|}{w/ SWD}  &  0.185 & 0.317 & 0.180\\
       \multicolumn{1}{|r|}{w/ MMD}  &  0.177 & 0.362 & 0.285 \\

       \hline
       Input + latent Reprogram & 0.176 & 0.373 & 0.197 \\
       \multicolumn{1}{|r|}{w/ SWD}  &  0.224 & 0.389 & 0.165\\
       \multicolumn{1}{|r|}{w/ MMD}  &  0.179 & 0.331 & 0.159\\
       
        \hline
    \end{tabular}
    \label{tab:asr}
    \vspace{-7mm}
\end{table}

\subsection{Experiment results}

We show the experiments on accent transferring from zh-CN to zh-TW. 
Table \ref{tab:tw} lists the results of MCD and human evaluations. We conduct the experiments on three PEL settings including adapter, input reprogram, and the combination of input reprogram and latent reprogram. Besides, we run the fine-tuning (FT) and decoder fine-tuning methods as baselines. By fine-tuning the whole backbone, we get the best value on MCD and MOS evaluations. Besides, we find out that all the methods we proposed shows competitive result against the full fine-tuning strategy and even outperform the decoder fine-tuning method. Compare to the result of parameter-efficient methods, the input reprogramming can have acceptable results With only 0.6\% of total parameters being trainable. The adapter and the joint reprogram (IR+LR) perform better results compared to the input-based methods but require more trainable parameters. We further show the results with auxiliary SWD/MMD loss to highlight the effect of optimal transport perspective. Clearly, the optimal transport viewpoint aids the model in producing natural speech with good accent quality.\footnote{Code and audio samples are available at \hyperlink{}{https://github.com/TTS-Research/PEL-TTS}} 

To show the effect of accent transfer, we produce synthetic Taiwanese Mandarin accent speech based on various PEL settings and compare the CER when testing on different pretrained ASR models. The results are shown in Table \ref{tab:asr}. Obviously, the synthetic speech cannot be well recognized by pretrained zh-CN model because of the large domain difference while performing in-domain testing with a pretrained zh-TW model achieves a lower CER value. One special finding is the fine-tuning method has a higher CER which indicates the sound with naturalness can't prove the ASR performance. Observing the CER on the zh-TW accent pretrained model, the error rate is reduced and the auxiliary MMD loss reaches a better result compared to SWD loss when utilizing latent parameter-efficient learning methods.  
\section{Conclusions}
We introduce parameter-efficient methods into text-to-speech accent adaptation via model reprogramming and residual adapter.  Benefits from input parameter-efficient learning, the input reprogramming, the backbone can repeatedly be deployed by only replacing the reprogramming layer. Furthermore, latent parameter-efficient learning including adapter learning and latent reprogramming shows their effect by tuning the latent feature, and improving the performance compare to the input reprogramming method. By leveraging the concept of optimal transport, we design an unsupervised auxiliary loss using SWD and MMD distance metrics to strengthen the tendency we observed from figure~\ref{fig:latent_distance}, and experiments prove that the auxiliary loss indeed helps the model produce speech with naturalness and higher accent similarity. Because we first introduce model reprogramming in text-to-speech and show its effectiveness on accent adaptation, it would be interesting to apply reprogramming schemes to other cases such as performing cross-lingual adaptation by leveraging the well-trained TTS model. \vspace{1mm}

\textbf{Acknowledgments}
The authors would like to express their gratitude to Heiga Zen and Bo Li from Google  for providing helpful insights and discussion on our draft.

\bibliographystyle{IEEEtran}
\bibliography{mybib}

\begin{thebibliography}{10}
\providecommand{\url}[1]{#1}
\csname url@samestyle\endcsname
\providecommand{\newblock}{\relax}
\providecommand{\bibinfo}[2]{#2}
\providecommand{\BIBentrySTDinterwordspacing}{\spaceskip=0pt\relax}
\providecommand{\BIBentryALTinterwordstretchfactor}{4}
\providecommand{\BIBentryALTinterwordspacing}{\spaceskip=\fontdimen2\font plus
\BIBentryALTinterwordstretchfactor\fontdimen3\font minus
  \fontdimen4\font\relax}
\providecommand{\BIBforeignlanguage}[2]{{%
\expandafter\ifx\csname l@#1\endcsname\relax
\typeout{** WARNING: IEEEtran.bst: No hyphenation pattern has been}%
\typeout{** loaded for the language `#1'. Using the pattern for}%
\typeout{** the default language instead.}%
\else
\language=\csname l@#1\endcsname
\fi
#2}}
\providecommand{\BIBdecl}{\relax}
\BIBdecl

\bibitem{zhang2023google}
Y.~Zhang, W.~Han \emph{et~al.}, ``Google usm: Scaling automatic speech
  recognition beyond 100 languages,'' \emph{arXiv preprint arXiv:2303.01037},
  2023.

\bibitem{zhang2022bigssl}
Y.~Zhang, D.~S. Park, W.~Han \emph{et~al.}, ``Bigssl: Exploring the frontier of
  large-scale semi-supervised learning for automatic speech recognition,''
  \emph{IEEE Journal of Selected Topics in Signal Processing}, vol.~16, no.~6,
  pp. 1519--1532, 2022.

\bibitem{brown2020language}
T.~Brown, B.~Mann \emph{et~al.}, ``Language models are few-shot learners,''
  \emph{NeurIPS}, vol.~33, pp. 1877--1901, 2020.

\bibitem{bommasani2021opportunities}
R.~Bommasani \emph{et~al.}, ``On the opportunities and risks of foundation
  models,'' \emph{arXiv preprint arXiv:2108.07258}, 2021.

\bibitem{popov2021grad}
V.~Popov, I.~Vovk, V.~Gogoryan \emph{et~al.}, ``Grad-tts: A diffusion
  probabilistic model for text-to-speech,'' in \emph{Proc. of ICML}, 2021, pp.
  8599--8608.

\bibitem{rombach2022high}
R.~Rombach, A.~Blattmann \emph{et~al.}, ``High-resolution image synthesis with
  latent diffusion models,'' in \emph{Proceedings of the IEEE/CVF Conference on
  Computer Vision and Pattern Recognition}, 2022, pp. 10\,684--10\,695.

\bibitem{ho2020denoising}
J.~Ho, A.~Jain, and P.~Abbeel, ``Denoising diffusion probabilistic models,''
  \emph{NeurIPS}, vol.~33, pp. 6840--6851, 2020.

\bibitem{lester2021power}
B.~Lester, R.~Al-Rfou, and N.~Constant, ``The power of scale for
  parameter-efficient prompt tuning,'' in \emph{Proc. of EMNLP}, 2021, pp.
  3045--3059.

\bibitem{rebuffi2017learning}
S.-A. Rebuffi, H.~Bilen, and A.~Vedaldi, ``Learning multiple visual domains
  with residual adapters,'' \emph{NeurIPS}, vol.~30, 2017.

\bibitem{elsayedadversarial}
G.~F. Elsayed, I.~Goodfellow, and J.~Sohl-Dickstein, ``Adversarial
  reprogramming of neural networks,'' in \emph{Proc. of ICLR}, 2019.

\bibitem{yang2021voice2series}
C.-H.~H. Yang, Y.-Y. Tsai, and P.-Y. Chen, ``{Voice2Series}: Reprogramming
  acoustic models for time series classification,'' in \emph{Proc. of ICML},
  2021, pp. 11\,808--11\,819.

\bibitem{hambardzumyan2021warp}
K.~Hambardzumyan, H.~Khachatrian, and J.~May, ``Warp: Word-level adversarial
  reprogramming,'' in \emph{Proc. of ACL}, 2021, pp. 4921--4933.

\bibitem{hetowards}
J.~He, C.~Zhou \emph{et~al.}, ``Towards a unified view of parameter-efficient
  transfer learning,'' in \emph{Proc. of ICLR}, 2023.

\bibitem{kumar2022fine}
A.~Kumar, A.~Raghunathan, R.~Jones, T.~Ma, and P.~Liang, ``Fine-tuning can
  distort pretrained features and underperform out-of-distribution,'' in
  \emph{Proc. of International Conference on Learning Representations}, 2022.

\bibitem{hung2023low}
Y.-N. Hung, C.-H.~H. Yang, P.-Y. Chen, and A.~Lerch, ``Low-resource music genre
  classification with cross-modal neural model reprogramming,'' in \emph{Proc.
  of ICASSP}.\hskip 1em plus 0.5em minus 0.4em\relax IEEE, 2023.

\bibitem{yang2023english}
C.-H.~H. Yang, B.~Li, Y.~Zhang \emph{et~al.}, ``From {E}nglish to more
  languages: Parameter-efficient model reprogramming for cross-lingual speech
  recognition,'' in \emph{Proc. of ICASSP}, 2023.

\bibitem{morioka2022residual}
N.~Morioka, H.~Zen, N.~Chen, Y.~Zhang, and Y.~Ding, ``Residual adapters for
  few-shot text-to-speech speaker adaptation,'' \emph{arXiv preprint
  arXiv:2210.15868}, 2022.

\bibitem{chien1999extraction}
J.-T. Chien, J.-C. Junqua, and P.~Gelin, ``Extraction of reliable
  transformation parameters for unsupervised speaker adaptation,'' in
  \emph{Sixth European Conference on Speech Communication and Technology},
  1999.

\bibitem{saon2012large}
G.~Saon and J.-T. Chien, ``Large-vocabulary continuous speech recognition
  systems: A look at some recent advances,'' \emph{IEEE signal processing
  magazine}, vol.~29, no.~6, pp. 18--33, 2012.

\bibitem{yamagishi2010thousands}
J.~Yamagishi \emph{et~al.}, ``Thousands of voices for hmm-based speech
  synthesis--analysis and application of tts systems built on various asr
  corpora,'' \emph{IEEE Transactions on Audio, Speech, and Language
  Processing}, vol.~18, no.~5, pp. 984--1004, 2010.

\bibitem{ji2004culture}
L.-J. Ji, Z.~Zhang, and R.~E. Nisbett, ``Is it culture or is it language?
  examination of language effects in cross-cultural research on
  categorization.'' \emph{Journal of personality and social psychology},
  vol.~87, no.~1, p.~57, 2004.

\bibitem{li-iscslp}
L.-J. Yang, I.-P. Yeh, and J.-T. Chien, ``Low-resource speech synthesis with
  speaker-aware embedding,'' in \emph{Proc. of International Symposium on
  Chinese Spoken Language Processing}, 2022, pp. 235--239.

\bibitem{tomanek2021residual}
K.~Tomanek, V.~Zayats, D.~Padfield, K.~Vaillancourt, and F.~Biadsy, ``Residual
  adapters for parameter-efficient asr adaptation to atypical and accented
  speech,'' in \emph{Proc. of EMNLP}, 2021, pp. 6751--6760.

\bibitem{xiao2023offsite}
G.~Xiao, J.~Lin, and S.~Han, ``Offsite-tuning: Transfer learning without full
  model,'' \emph{arXiv preprint arXiv:2302.04870}, 2023.

\bibitem{chenadaspeech}
M.~Chen, X.~Tan, B.~Li, Y.~Liu, T.~Qin, T.-Y. Liu \emph{et~al.}, ``{AdaSpeech}:
  Adaptive text to speech for custom voice,'' in \emph{Proc of ICLR}, 2021.

\bibitem{gu2021ppt}
Y.~Gu, X.~Han, Z.~Liu, and M.~Huang, ``{PPT}: Pre-trained prompt tuning for
  few-shot learning,'' in \emph{Proc. of ACL}, 2022, pp. 8410--8423.

\bibitem{houlsby2019parameter}
N.~Houlsby, A.~Giurgiu \emph{et~al.}, ``Parameter-efficient transfer learning
  for {NLP},'' in \emph{Proc. of ICML}, 2019, pp. 2790--2799.

\bibitem{sung2022vl}
Y.-L. Sung, J.~Cho, and M.~Bansal, ``{VL}-adapter: Parameter-efficient transfer
  learning for vision-and-language tasks,'' in \emph{Proc. of CVPR}, 2022, pp.
  5227--5237.

\bibitem{renfastspeech}
Y.~Ren, C.~Hu \emph{et~al.}, ``Fastspeech 2: Fast and high-quality end-to-end
  text to speech,'' in \emph{Proc. of ICLR}, 2018.

\bibitem{watanabe2018espnet}
S.~Watanabe and T.~Hori~\textit{et al.}, ``Espnet: End-to-end speech processing
  toolkit,'' \emph{INTERSPEECH}, 2018.

\bibitem{snyder2018x}
D.~Snyder, D.~Garcia-Romero \emph{et~al.}, ``X-vectors: Robust dnn embeddings
  for speaker recognition,'' in \emph{Proc. of ICASSP}.\hskip 1em plus 0.5em
  minus 0.4em\relax IEEE, 2018, pp. 5329--5333.

\bibitem{chang2022exploration}
K.-W. Chang, W.-C. Tseng \emph{et~al.}, ``An exploration of prompt tuning on
  generative spoken language model for speech processing tasks,'' \emph{Proc.
  Interspeech}, 2022.

\bibitem{radford2021learning}
A.~Radford, J.~W. Kim \emph{et~al.}, ``Learning transferable visual models from
  natural language supervision,'' in \emph{Proc. of ICML}, 2021, pp.
  8748--8763.

\bibitem{baevski2020wav2vec}
A.~Baevski, Y.~Zhou, A.~Mohamed, and M.~Auli, ``wav2vec 2.0: A framework for
  self-supervised learning of speech representations,'' \emph{Prof. of
  NeurIPS}, vol.~33, pp. 12\,449--12\,460, 2020.

\bibitem{peyre2019computational}
G.~Peyr{\'e}, M.~Cuturi \emph{et~al.}, ``Computational optimal transport: With
  applications to data science,'' \emph{Foundations and Trends{\textregistered}
  in Machine Learning}, vol.~11, no. 5-6, pp. 355--607, 2019.

\bibitem{scholkopf2002learning}
B.~Sch{\"o}lkopf, A.~J. Smola, F.~Bach \emph{et~al.}, \emph{Learning with
  kernels: support vector machines, regularization, optimization, and
  beyond}.\hskip 1em plus 0.5em minus 0.4em\relax MIT press, 2002.

\bibitem{chien2015hierarchical}
J.-T. Chien, ``Hierarchical pitman--yor--dirichlet language model,''
  \emph{IEEE/ACM Transactions on Audio, Speech, and Language Processing},
  vol.~23, no.~8, pp. 1259--1272, 2015.

\bibitem{feydy2019interpolating}
J.~Feydy \emph{et~al.}, ``Interpolating between optimal transport and mmd using
  sinkhorn divergences,'' in \emph{AISTATS}.\hskip 1em plus 0.5em minus
  0.4em\relax PMLR, 2019, pp. 2681--2690.

\bibitem{yang2021multi}
C.-H.~H. Yang, L.~Liu, Y.~Gu \emph{et~al.}, ``Multi-task language modeling for
  improving speech recognition of rare words,'' in \emph{Proc. of ASRU}.\hskip
  1em plus 0.5em minus 0.4em\relax IEEE, 2021, pp. 1087--1093.

\bibitem{aishell}
Y.~Shi, H.~Bu, X.~Xu, S.~Zhang, and M.~Li, ``Aishell-3: A multi-speaker
  mandarin tts corpus and the baselines,'' \emph{Proc. of Interspeech}, 2020.

\bibitem{xvectors}
D.~Snyder, D.~Garcia-Romero \emph{et~al.}, ``X-vectors: Robust dnn embeddings
  for speaker recognition,'' in \emph{Proc. of ICASSP}, 2018, pp. 5329--5333.

\bibitem{yamamoto2020parallel}
R.~Yamamoto, E.~Song, and J.-M. Kim, ``Parallel wavegan: A fast waveform
  generation model based on generative adversarial networks with
  multi-resolution spectrogram,'' in \emph{Proc. ICASSP}, 2020, pp. 6199--6203.

\bibitem{ardila2020common}
R.~Ardila, Branson \emph{et~al.}, ``Common voice: A massively-multilingual
  speech corpus,'' in \emph{Proceedings of the 12th Language Resources and
  Evaluation Conference}, 2020, pp. 4218--4222.

\bibitem{finkelstein2022training}
L.~Finkelstein, H.~Zen \emph{et~al.}, ``Training text-to-speech systems from
  synthetic data: A practical approach for accent transfer tasks,'' \emph{arXiv
  preprint arXiv:2208.13183}, 2022.

\end{thebibliography}

\end{document}